# Identifying the Multimodal Hierarchy of Public Transit Systems Using Itinerary Data

**Junhee Lee[1], Seungmo Kang[2,*], and Jinwoo Lee[2,*]**

1: Department of Civil, Environmental and Architectural Engineering, Korea University, Seoul, Republic of Korea, 02841; ORCID: 0009-0008-7001-6151
2: School of Civil, Environmental and Architectural Engineering, Korea University, Seoul, Republic of Korea, 02841; ORCID: 0000-0002-9435-5835
3: Department of Civil and Environmental Engineering, Korea Advanced Institute of Science and Technology, Daejeon, Korea, 34141; ORCID: 0000-0002-6692-9715
*: Co-Corresponding Authors; s_kang@korea.ac.kr and jinwoo@kaist.ac.kr

**ABSTRACT**
As urban mobility integrates traditional and emerging modes, public transit systems are becoming increasingly complex. Some modes complement each other, while others compete, influencing users' multimodal itineraries. To provide a clear, high-level understanding of these interactions, we introduce the concept of a macroscopic multimodal hierarchy. In this framework, trips follow an "ascending-descending" order, starting and ending with lower hierarchical modes (e.g., walking) that offer high accessibility, while utilizing higher modes (e.g., subways) for greater efficiency. We propose a methodology to identify the multimodal hierarchy of a city using multimodal smart card itinerary data and demonstrate its application with actual data collected from Seoul and the surrounding metropolitan area in South Korea.

## 1. INTRODUCTION

Effective public transit planning requires a clear understanding of the hierarchy of transportation modes (Wang et al., 2020). Passengers often categorize these modes into a hierarchy, such as walking, feeder lines, and trunk lines, with each serving a different purpose. By incorporating this hierarchy into transit network design, planners can better organize routes, ensuring that lower-tier modes act as feeders to higher-tier modes in a specific area.

Although transit hierarchies are typically based on factors like efficiency and accessibility, there has been limited research on whether the hierarchy designed by planners matches how passengers actually use the system. This gap presents an opportunity to examine real-world travel patterns and uncover any discrepancies using smart card data. By analyzing actual passenger usage, we can get a clearer picture of how the transit hierarchy functions in practice.

This study aims to address this gap by introducing a new method for analyzing the hierarchy of public transit modes based on real passenger travel patterns. Using smart card data, we demonstrate this approach with data from the Seoul metropolitan area. Our method classifies the first and last parts of a trip as the "ascending" and "descending" stages, respectively, based on the total travel distance from the starting point to the final destination. We then analyze passenger transfers between transit modes to uncover hierarchical relationships.

Previous studies, such as Daganzo and Ouyang (2019) and Oh et al. (2024), have shown that multi-modal trips typically follow a low → high → low hierarchy: passengers first move from low-level modes (low efficiency, high accessibility) to high-level modes (high efficiency, low accessibility), and then return to lower-level modes near their destination. However, these studies assume spatial homogeneity, which can overlook the role of localized transit modes like light rail transit (LRT) or community buses. These modes often operate within smaller areas and may not fit neatly into the assumed hierarchy, making it difficult to accurately capture their role in the broader transit system. Our proposed method addresses this limitation by not assuming spatial homogeneity, allowing for a more nuanced and accurate identification of the modal hierarchy.

## 2. METHODOLOGY

A target multi-modal urban transportation system involves the set of modes, $\mathcal{M} = \{1, \dots, M\}$, where the total number of modes is $M$. The objective is to identify the macroscopic hierarchy among $M$ modes. To achieve this, the ascending-descending method classifies transfers based on half of the total travel distance. Using this classification, the multimodal transfer rate matrix is constructed to analyze directional transfer tendencies, and the multimodal hierarchy distance quantifies hierarchical differences between modes. The average hierarchical order of each mode is then computed and rescaled for comparison, where higher values indicate superior hierarchy. Finally, in multi-zonal systems, interzonal and intrazonal hierarchical structures are analyzed separately using corresponding itineraries.

### 2.1 Single-Zonal Hierarchy Analysis

We consider a set of $N$ multi-modal itineraries, $\mathcal{N} = \{1, \dots, N\}$, each associated with a corresponding O-D pair. Walk-only trips are excluded from $\mathcal{N}$, and itinerary $n \in \mathcal{N}$ comprises of one or more non-walking modes, such as bus or metro. The first and last sub-trips in every itinerary are always assumed to be walking trips. The stages of public transportation use by passengers are categorized into two phases: the ascending phase and the descending phase. These phases are distinguished based on the total travel distance ($L^n$) from the initial boarding point to the final alighting point for each passenger. The ascending and descending phases are determined based on half of $L^n$. Transfers occurring before reaching half of $L^n$ are classified as ascending-phase transfers, whereas those occurring after this point are classified as descending-phase transfers. The total numbers of ascending and descending transfers from mode $i$ to mode $j$ for itinerary $n$ are denoted by $a_{ij}^n$ and $d_{ij}^n$, respectively.

To analyze the predominant direction of intermodal transfers occurring at each phase, the following matrix was constructed. Based on the definition of $a_{ij}^n$, summing over all $n$ yields the total number of transfers from mode $i$ to mode $j$ during the ascending phase, denoted as $a_{ij}$, as presented in (1). Similarly, summing $d_{ij}^n$ over all $n$ provides the total number of transfers from mode $i$ to mode $j$ during the descending phase, denoted as $d_{ij}$, as shown in (2).

$$a_{ij} = \sum_{\forall n} a_{ij}^n, \forall i, j \in \mathcal{M} \tag{1}$$

$$d_{ij} = \sum_{\forall n} d_{ij}^n, \forall i, j \in \mathcal{M} \tag{2}$$

We define the transfer rate matrix for the ascending phase as $A$ and the transfer rate matrix for the descending phase as $D$, where each element in row $i$ and column $j$ is denoted as $A_{ij}$ and $D_{ij}$, respectively, as follows:

$$A_{ij} = \frac{a_{ij}}{a_{ij} + a_{ji}}, \forall i, j \in \mathcal{M} \tag{3}$$

$$D_{ij} = \frac{d_{ij}}{d_{ij} + d_{ji}}, \forall i, j \in \mathcal{M} \tag{4}$$

To quantify the hierarchical difference between modes identified through the transfer rate matrices for each phase, the multimodal hierarchy distance is defined as the difference in transfer rates between two modes. Specifically, the hierarchy distance of mode $i$ relative to mode $j$ in the ascending phase is denoted as $A_{ij}^*$, while the hierarchy distance of mode $i$ relative to mode $j$ in the descending phase is denoted as $D_{ij}^*$, as follows:

$$A_{ij}^* = A_{ij} - A_{ji}, \in [-1,1] \tag{5}$$

$$D_{ij}^* = D_{ji} - D_{ij}, \in [-1,1] \tag{6}$$

Using the previously defined $A_{ij}^*$ and $D_{ij}^*$, we compute the average hierarchical distance of mode $i$ relative to all other modes. The average hierarchical distance for mode $i$ in the ascending and descending phases is denoted as $A_i^*$ and $D_i^*$, respectively. These values are rescaled from the original range of $[-1,1]$ to $[0,1]$ to facilitate comparison. A value of $A_i^*$ or $D_i^*$ close to 1 indicates a higher hierarchical status.

$$A_i^* = \frac{1}{2}\left(\sum_{j \in \mathcal{M}/\{i\}} \frac{A_{ij}^*}{M - 1}\right) + \frac{1}{2}, \forall i \in \mathcal{M} \tag{7}$$

$$D_i^* = \frac{1}{2}\left(\sum_{j \in \mathcal{M}/\{i\}} \frac{D_{ij}^*}{M - 1}\right) + \frac{1}{2}, \forall i \in \mathcal{M} \tag{8}$$

The overall hierarchy of mode $i$, denoted as $H_i$, is defined as the average hierarchical distance of mode $i$ relative to all other modes. The rescaled hierarchical value, $H_i$, is normalized within the range $[0,1]$ and is defined as:

$$H_i^* = \frac{A_i^* + D_j^*}{2}, \forall i \in \mathcal{M} \tag{9}$$

If $H_i^* < H_j^*$, mode $j$ is macroscopically superior to mode $i$ in terms of hierarchical structure.

### 2.2 Multi-Zonal Hierarchy Analysis

If the system under consideration consists of multiple distinct zones, where the set of zones is denoted as $\mathcal{P}$, a multi-zonal analysis can be conducted to examine the hierarchical structure across different zones. To obtain interzonal multimodal hierarchical information, we use the set of interzonal itineraries that originate from zone $p \in \mathcal{P}$ and are destined for zone $q \in \mathcal{P}/\{p\}$, denoted as $\mathcal{N}^{pq}$,. The hierarchical measures for interzonal trips—namely, $A_i^{pq*}$, $D_i^{pq*}$, and $H_i^{pq*}$—are computed in the same manner as $A_i^*$, $D_i^*$, and $H_i^*$ using $\mathcal{N}$ in the previous subsections. The intrazonal hierarchical information for zone $p \in \mathcal{P}$, denoted as $A_i^{pp*}$, $D_i^{pp*}$, and $H_i^{pp*}$, is obtained using the set of intrazonal itineraries, $\mathcal{N}^{pp}$.

## 3. CASE STUDY

We use the T-money transit card dataset collected on November 12, 2019, for the Seoul Metropolitan Area including Seoul City, Gyeonggi Province, and Incheon City, South Korea. The dataset contains information such as boarding and alighting station IDs, transport modes, travel distances, and travel times. The final dataset contained a total of 10,264,700 itineraries ($N$) meeting our flitering criteria. The distribution of these itineraries is as follows: Itineraries corresponded to travel within Seoul ($N^{11}$) is 7,500,141; Itineraries corresponded to travel from Seoul to Gyeonggi or Incheon ($N^{12}$) is 1,365,156; Itineraries corresponded to travel from Gyeonggi or Incheon to Seoul ($N^{21}$) is 1,399,403. The modes considered in this analysis include walking ($i = 1$), community bus ($i = 2$), urban bus ($i = 3$), intercity bus ($i = 4$), light rail ($i = 5$), and metro ($i = 6$), i.e., $\mathcal{M} = \{1, \dots, 6\}$ with $M = 6$. The spatial layout of them is presented in **Figure 1**. Additional details about the dataset are omitted in this extended abstract, which can be provided upon request.

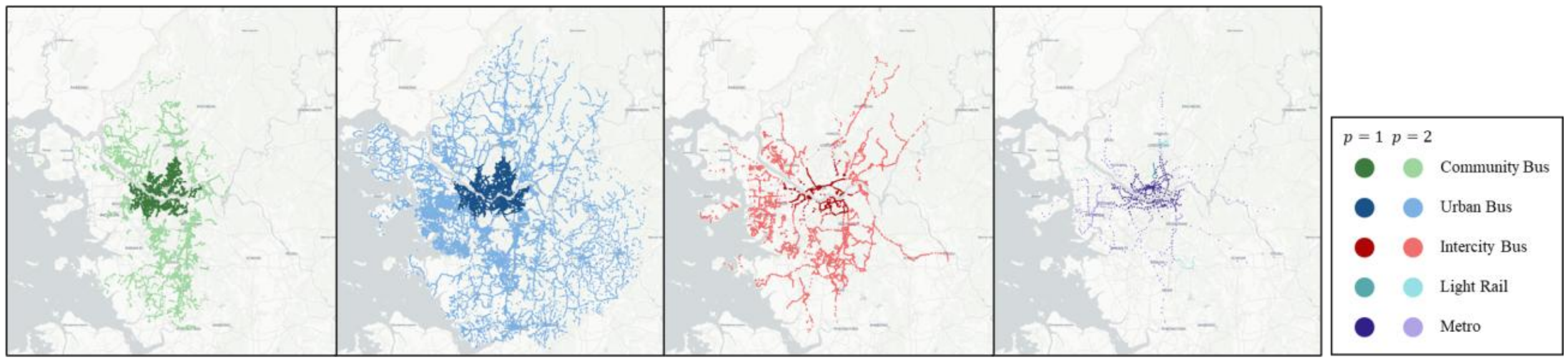

**Figure 1** Spatial Distribution of Stops by Mode in the Seoul Metropolitan Area

We first examine the multimodal hierarchy within Seoul ($p = 1$) using only intrazonal Itineraries. The details of the average hierarchical order are summarized in **Figure 2**. The orange diagonal line in this figure indicates the points where $A_i^{11*}$ and $D_i^{11*}$ are equal. The closer a mode's average hierarchical order is to this line, the smaller the difference between its hierarchical order in the ascending and descending phases, implying minimal variation in hierarchy between these two phases. As shown in **Figure 2**, a comparison of the average hierarchical order between the ascending and descending phases revealed no significant differences across all modes, i.e., $A_i^{11*} \approx D_i^{11*} \approx H_i^{11*}$. The hierarchical order of each mode remained consistent across both phases, following the sequence of walking, community bus, urban bus, light rail, and metro.

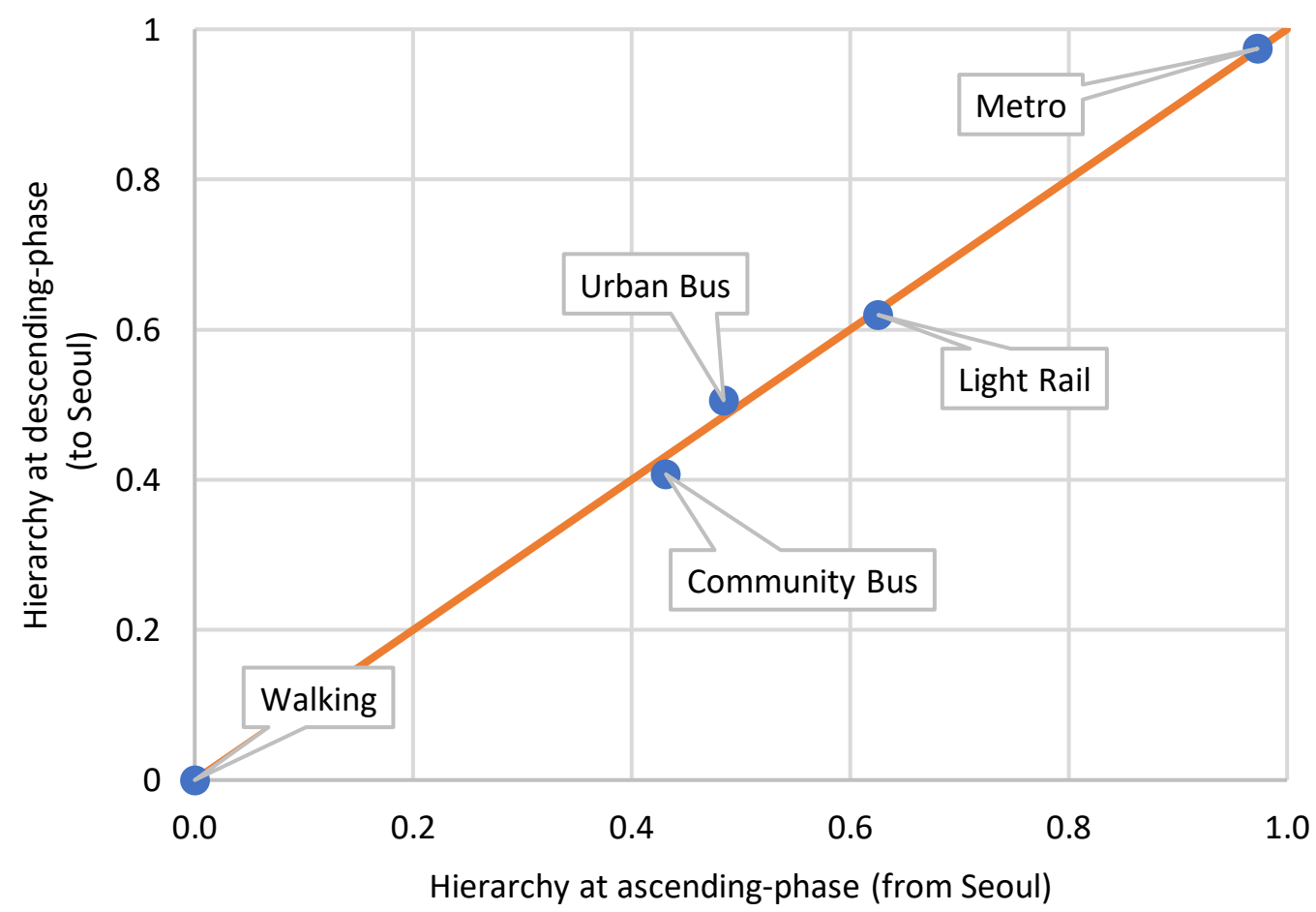

**Figure 2** Comparison of hierarchical levels by mode and phase in $\mathcal{N}^{11}$

Second, we analyze the hierarchy between Seoul ($p = 1$) and Gyeonggi and Incheon ($p = 2$) using interzonal Itineraries. The results of the average hierarchical order for $\mathcal{N}^{12}$ and $\mathcal{N}^{21}$ are summarized in **Figure 3a** and **Figure 3b**, respectively. The lowest hierarchical order was assigned to the community bus, followed by the urban bus, the light rail, the metro, and finally the intercity bus, which had the highest hierarchical order. This result aligns with the hierarchy observed in $\mathcal{N}^{11}$, where the intercity bus was excluded from the analysis. However, unlike in $\mathcal{N}^{11}$, the values of $H_i^{12*}$ and $H_i^{21*}$ for the community bus, urban bus, and metro were relatively lower compared to $H_i^{11*}$. This suggests that the inclusion of the intercity bus in $\mathcal{N}^{12}$ and $\mathcal{N}^{21}$ led to a relative decrease in the average hierarchical order of other modes. The observed variation in average hierarchical order depending on the modes considered indicates that the relative differences between transportation modes can be effectively assessed through this method.

We further analyzed the modal hierarchy across different zones using the hierarchical structures derived from $A_i^{12*}$, $D_i^{12*}$, $A_i^{21*}$, and $D_i^{21*}$. The results indicate that the hierarchical order of the intercity bus and the metro differs between Seoul and Gyeonggi and Incheon, reflecting a regional variation in transportation hierarchy. This is because in Seoul, metro stations are more densely distributed than intercity bus stops, whereas in Gyeonggi and Incheon, intercity bus stops are more prevalent than metro stations. Additionally, we observed a difference in the hierarchical position of the urban bus between Seoul ($p = 1$) and Gyeonggi and Incheon ($p = 2$), which indicates that the hierarchical order of urban bus in Gyeonggi and Incheon holds a relatively lower hierarchical position compared to that in Seoul. These findings confirm that the average hierarchical order of a given mode can vary depending on the region, reflecting differences in the relative position of transportation modes across different areas.

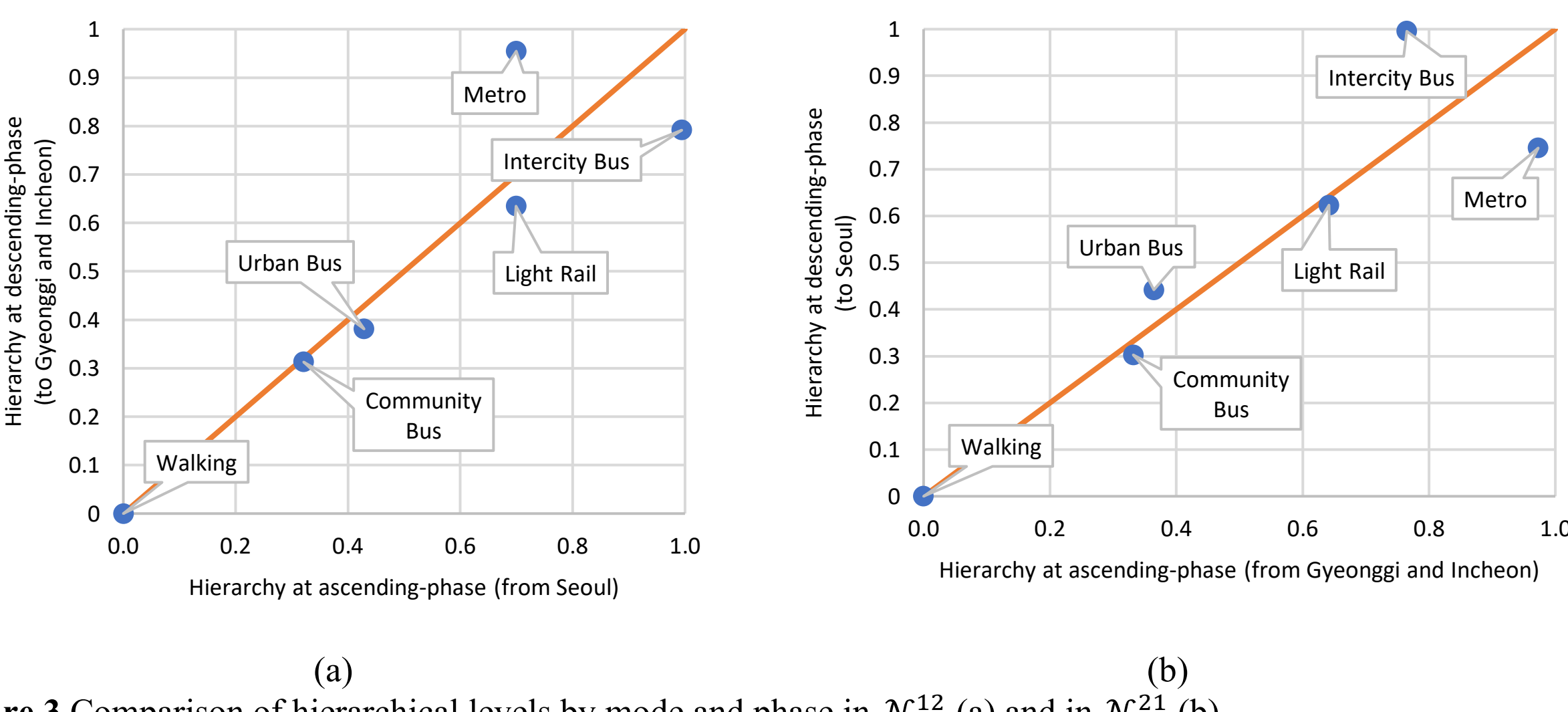

(a) (b)

**Figure 3** Comparison of hierarchical levels by mode and phase in $\mathcal{N}^{12}$ (a) and in $\mathcal{N}^{21}$ (b)

## ACKNOWLEDGEMENTS

This paper has not been published and is currently under review.